\documentclass[%
 aip,
 rsi,
 amsmath,amssymb,
 preprint,%
]{revtex4-1}

\usepackage{graphicx}
\usepackage{dcolumn}
\usepackage{bm}

\usepackage[utf8]{inputenc}
\usepackage[T1]{fontenc}
\usepackage{mathptmx}
\usepackage{etoolbox}

\makeatletter
\def\@email#1#2{%
 \endgroup
 \patchcmd{\titleblock@produce}
  {\frontmatter@RRAPformat}
  {\frontmatter@RRAPformat{\produce@RRAP{*#1\href{mailto:#2}{#2}}}\frontmatter@RRAPformat}
  {}{}
}%
\makeatother
\begin{document}

\preprint{AIP/123-QED}

\title[]{Thomson scattering diagnostic at the gas dynamic trap}
\author{A. Lizunov}
\email{lizunov@inp.nsk.su}
\author{T. Berbasova}
\author{A. Khilchenko}
\author{A. Kvashnin}
\author{E. Puryga}
\affiliation{Budker Institute of nuclear physics, 630090 Novosibirsk, Russia}
\author{A. Sandomirsky}
\affiliation{Novosibirsk State University, 630090 Novosibirsk, Russia}
\author{P. Zubarev}
\affiliation{Budker Institute of nuclear physics, 630090 Novosibirsk, Russia}

\date{\today}

\begin{abstract}
The incoherent Thomson scattering diagnostic with multiple line of sight is installed at the gas dynamic trap (GDT) for measurements of radial profiles
of the plasma electron temperature and density. The diagnostic is based on a Nd:YAG laser with the wavelength of $1064~nm$. Its collecting lens uses $90^\circ$ scattering geometry having eleven lines of sight in total. At the current stage, six of them are equipped with interference filter spectrometers. Results of the initial operation within a month-long experimental campaign at GDT show that radial profiles are delivered with the accuracy of few percent for $T_e \geqslant 20~eV$ 
and $n_e \geqslant 10^{19}~m^{-3}$. The next planned boost of the diagnostic performance is upgrade with the pulse burst laser to study a fast dynamics of plasma parameters in every GDT shot.
\end{abstract}

\maketitle

\section{\label{sec:intro} Introduction}
The gas dynamic trap (GDT) is a linear mirror magnetic system for confinement of plasmas with energetic ions created by injection of high-power atomic beams \cite{gdt-review-ppcf}. A gradual progress in many-year experiments backed by solid theoretical explanations has been shown that a relatively
high electron temperature of the central cell plasma $T_e \simeq 1~keV$ is achievable \cite{ecrh_prl} despite the open magnetic field geometry where field lines meet end plates in the regions beyond mirrors. Such regimes of a decent electron temperature and an extreme $\beta \sim 0.5$ can be enabled \cite{gdt_maxbeta} thanks to the ambipolar potential barrier developing in the expander region \cite{axial_conf, gdt_ampotential} and so called vortex stabilization of magnetohydrodynamic (MHD) modes of instability \cite{gdt_vortex-1}. Since these findings are validated, there are several ways of evolution of the GDT concept  being investigated. A long range sight is aimed at a large scale research facility serving as an intermediate step towards a fusion reactor \cite{gdmt, gdt_dia_trap-1}. The most promising and well established near-term perspective is a high luminosity source of 14-MeV neutrons \cite{gdt_ns-1} for material studies and other applications. Besides the gas dynamic trap, a significant success in plasma confinement and heating in the related linear trap C-2U is demonstrated  in the TAE Technologies company as well \cite{tae_Gota_2017}.

The paper describes the recently developed incoherent Thomson scattering (TS) diagnostics for GDT for the measurement of spatial distributions of the plasma electron temperature and density. This TS system is believed to be an important new instrument of the GDT diagnostics array for all the upcoming research activity.  
The paper is structured in the following way: the Section~\ref{sec:opsys} describes the construction, laser beamline and light collection optical systems; the Section~\ref{sec:codaq} covers control and data acquisition and processing topics; the Section~\ref{sec:results} discusses the method of calculation and presents results of measurements. Finally, the Section~\ref{sec:summary} addresses the questions of diagnostic performance and future plans. 

\section{\label{sec:opsys} Construction and optical system of TS diagnostic}
\subsection{\label{sec:beamline}Beamline}
\begin{figure}[htbp]
\centering 
\includegraphics[width=1.\textwidth]{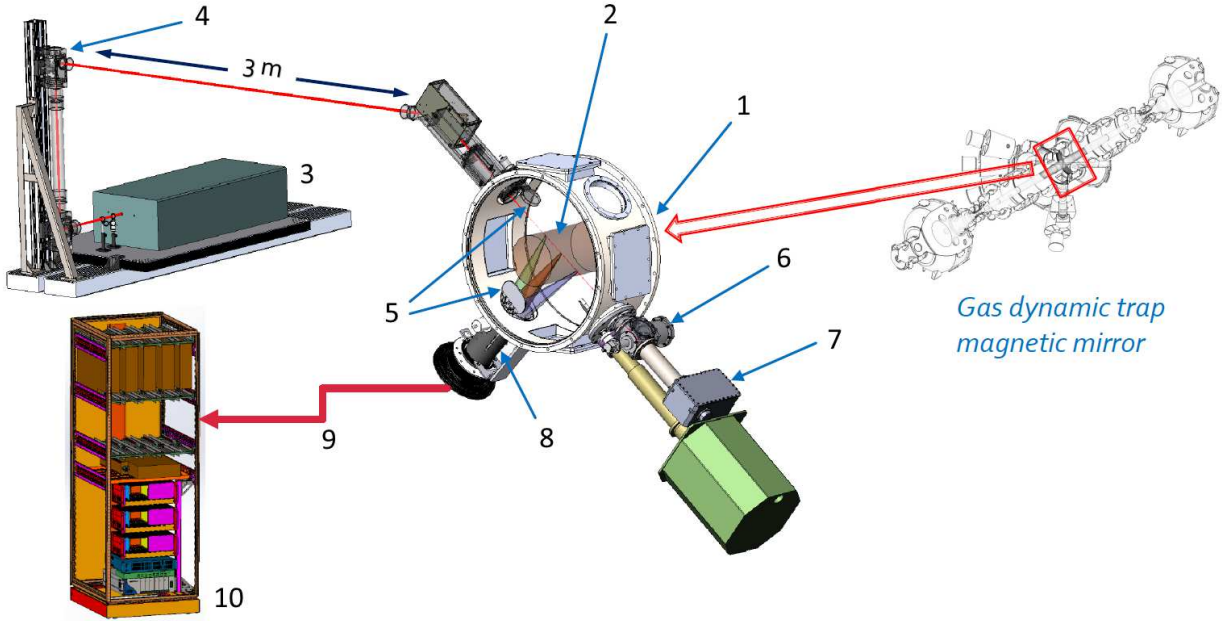}
\caption{\label{fig:layout} Layout of the TS diagnostic at GDT: 1 -- central section of the GDT vacuum vessel (scaled part inside the red rectangle), 2 -- plasma, 3 -- optical breadboard with the Nd:YAG laser, 4 -- motorized alignment unit, 5 -- protection shutters for optical ports, 6 -- alignment red beam monitor, 7 -- beam dump, 8 -- optical system for light collection, 9 -- optical fibre assemblies, 10 -- diagnostic cabinet with spectrometers.}
\end{figure}

\begin{figure}[htbp]
\centering 
\includegraphics[width=1.\textwidth]{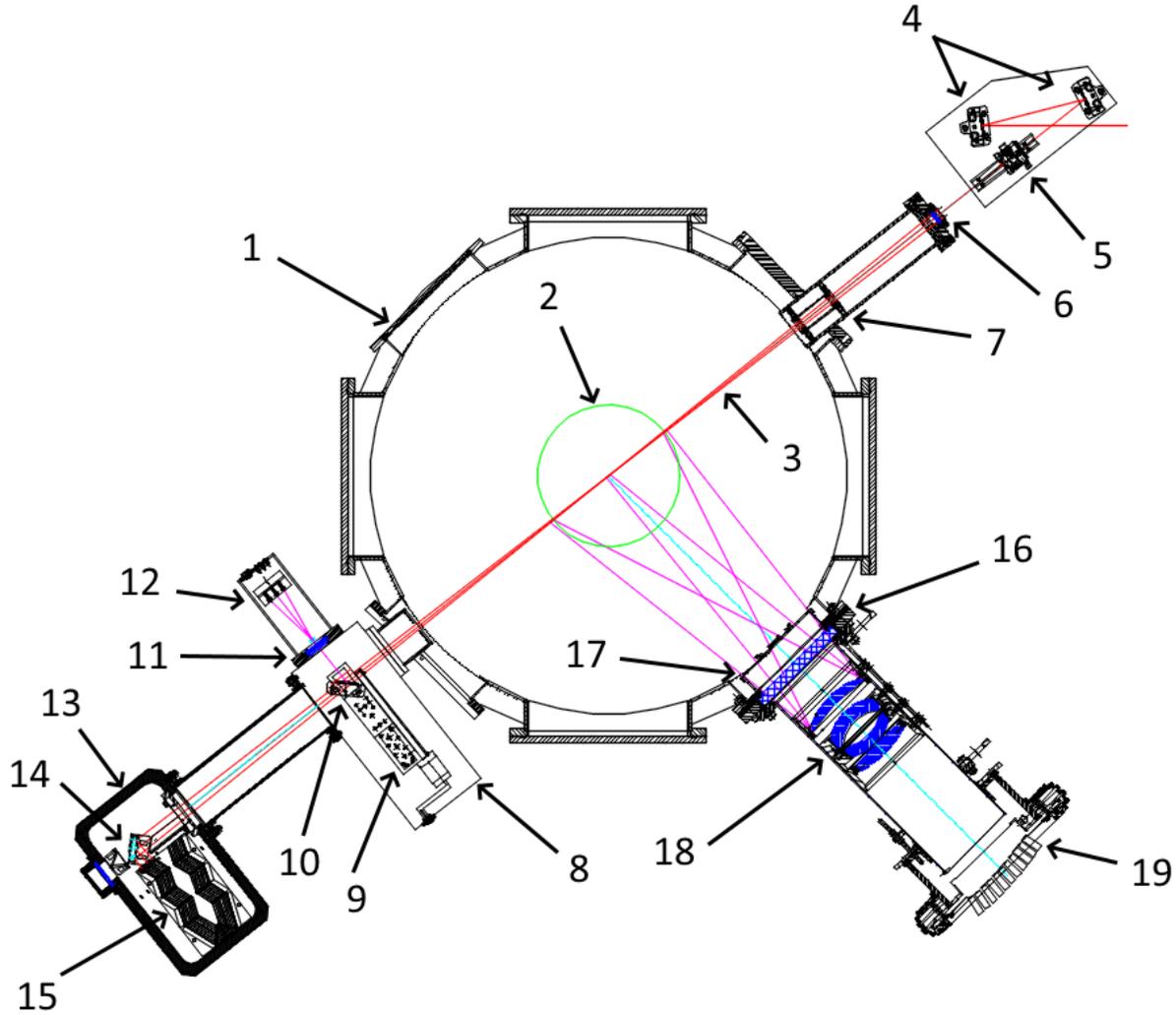}
\caption{\label{fig:geom} Beamline and scattering geometry : 1 -- midplane cross section of the GDT vacuum vessel, 2 -- plasma ($r=15.5~cm$), 3 -- laser beam focused on the machine axis, 4 -- flat dielectric mirrors, 5 -- AR-coated PCX lens $f=100~cm$, 6 -- AR-coated N-BK7 vacuum window at $0^\circ$ incidence,  7 -- set of input apertures, 8 -- vacuum volume of the alignment monitor, 9 -- linear translation stage with the stepper motor, 10 -- $45^\circ$ mirror for the alignment red beam, 11 -- output vacuum window and plano-concave lens for the red beam, 12 -- 5 photodiodes FDS100 arranged in a cross pattern, 13 -- beam dump vacuum volume, 14 -- $45^\circ$ dielectric mirror for 1064~nm, 15 -- beam absorption labyrinth of folded sheets of 0.2~mm Mo, 16 -- diagnostic port of TS collecting optics with the vacuum window (AR-coated N-BK7), 17 -- protecting shutter, 18 -- six-element achromatic lens, 19 -- fibre optical assembly collectors arranged along the image surface.}
\end{figure}

The general diagnostic layout is shown if Fig.~\ref{fig:layout}.
All the TS components are arranged in the GDT experimental hall. The diagnostic has a single Nd:YAG laser (Spectra Physics model Quanta Ray Pro 250) at the fundamental harmonic with the wavelength $\lambda=1064~nm$ and the optical registration system using passband interference filters for spectral sensitivity. Main  parameters are listed in the Table~\ref{tab:opsys_param}. The laser operates at a constant pulse repetition frequency of 10~Hz. This gives a single time point within the pulsed GDT plasma experiment, which has the heating phase duration of $\approx 10~ms$. Plans to progress towards time resolved measurements are discussed in the Section~\ref{sec:summary}. The beamline bends as shown in Fig.~\ref{fig:layout}, where the one of  $45^\circ$ hard-coated dielectric mirrors passes the red alignment beam along the Nd:YAG beam path. The motorized angular stage on top of the vertical stand (item 4 in Fig.~\ref{fig:layout}) serves for a remote control over the beamline alignment. The total beamline length is 8~m. Combined probing and alignment beams enter the GDT vacuum vessel at the central plane. Fig.~\ref{fig:geom} shows the outline of the Thomson scattering geometry.
\begin{table}[htbp]
\caption{\label{tab:opsys_param} Main parameters of the TS optical system}
\begin{ruledtabular}
\begin{tabular}{ll}
\it{Laser} & \\
\hline
Wavelength & $1064~nm$ \\
Pulse energy & $1.5~J$ \\
Pulse duration & $10~ns$ \\
Repetition frequency & $10~Hz$ \\
Beam diameter & $8~mm$ \\
Angular divergence & $1.5~mrad$ \\
\hline
\it{Optical system} & \\
\hline
Total number of LOS & 11\footnote{In use at the moment: 6.} \\
$f/ \#$  & 3.1 \\
Scattering angle & $83.3^\circ \div 109.7^\circ$ \\
Observation points along LOS & $-15.6 \div 15.2~cm$ \\
Resolution $\delta r \times \delta z$ & $10 \times 1~mm$ \\
\end{tabular}
\end{ruledtabular}
\end{table}

The beam entrance has a N-BK7 glass vacuum window with the anti-reflection (AR) coating at $0^\circ$ incidence and a set of apertures on the vacuum side intercepting the stay laser radiation from window and reducing the net backlight level. A single plano-convex (PCX) lens $f=100~cm$ focuses the beam onto the machine axis. Beam ray tracing in ZEMAX confirmed by measurements, gives the spot sizes of $\delta r_{centre} \cong 0.7~mm$ and $\delta r_{edge} \cong 3.2~mm$, for the on-axis position and the imaging plasma edge radius of $r_{edge}=15.5~cm$, respectively. Considering the collecting lens magnification of $M \cong 0.6$, above $99 \%$ of the image fits the fibre optical input strip for the on-axis point. This power fraction amounts $\approx 0.7$ for the edge point. The output laser beamline is arranged completely inside the vacuum volume without any output windows. On the GDT chamber output, a movable $45^\circ$ mirror mounted on the linear translation stage powered by a stepper motor. This drive allows for a precise and reproducible mirror positioning and reflection of the red beam towards the five FDS100 photodiodes (see 8...12 in Fig.~\ref{fig:geom}). The dedicated software extracts the actual beam spot displacement comparing to the best alignment reference and adjusts the mirror~4 (see Fig.~\ref{fig:layout}) via the feedback to eliminate the shift. This alignment check procedure is presumed to make once a workday. In this mode, the Nd:YAG beam on the laser breadboard is diverted from the beamline and the movable mirror 10 is set to the position in the red ray as illustrated in Fig.~\ref{fig:geom}. For TS measurements (normal working mode), the translation stage parks this mirror at the farthest position of $\approx 15~cm$ outside the beam. We observed that the red beam spot on diodes was stable within $1~mm$ during a month of experiments so no  alignment correction is necessary so far. The described setup will be useful in a long term perspective as a technique for a routine {\it in-situ} alignment monitoring.
\subsection{\label{sec:light_collection}Collecting lens}
The light collecting optical system has 11 observation lines in total, see Fig.~\ref{fig:geom} and Table~\ref{tab:opsys_param}. In the present setup, 6 filter spectrometers \cite{TS_GDT-spectro} are coupled via 10~m long fibre optical bundles with LOS distributed over the plasma radius. The lens has the diameter of $160~mm$, which is the maximum allowed by mechanical restrictions of GDT diagnostic ports, see Fig.\ref{fig:geom}. This size gives the throughput of $f/3.1$ (numerical aperture $NA \cong 0.16$) agreed with apertures of the optical fibre and the spectrometer. To ease the alignment with a visible red beam $@635~nm$, the achromatic scheme is used in the six-element collecting lens design having the residual chromatic focal shift less than $0.15~mm$. The point spread function holds below 0.43~mm across the image surface.

Spectrometers are mounted in the diagnostic cabinet among other hardware of the control and data acquisition system, refer to the Section~\ref{sec:codaq} for details. The noted above paper \cite{TS_GDT-spectro} gives a brief comparison of the spectrometer made for the GDT TS with similar instruments.

\section{\label{sec:codaq} Control and data acquisition system}
\begin{figure}[htbp]
\centering 
\includegraphics[width=1.\textwidth]{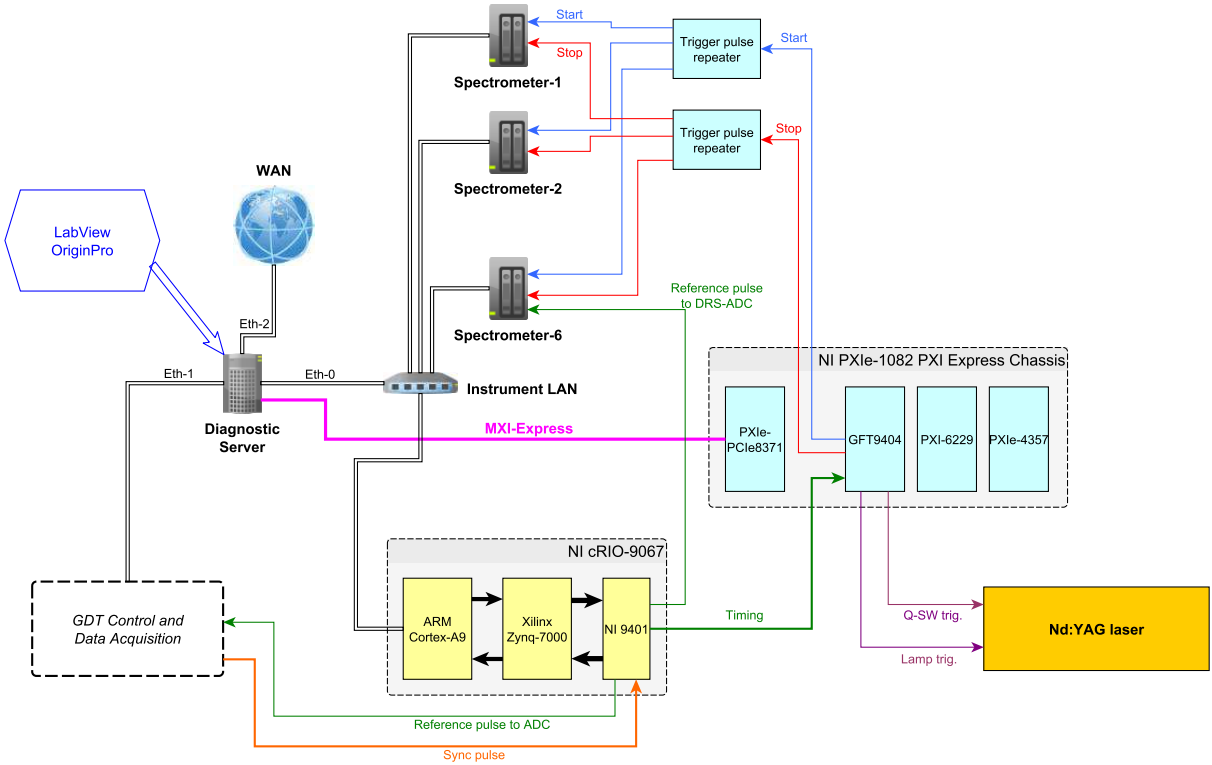}
\caption{\label{fig:timing} Block diagram of the TS timing system.}
\end{figure}

TS diagnostic is an autonomous subsystem fed from a separate uninterrupted 3-phase power supply. Besides optomechanical structures, it has the set of electronic equipment and its own data management and storage system. Communications between TS and common GDT CODAQ comprise the external sync pulse and streaming of processed physical data to the GDT archive via the TCP. The control system hardware is partially distributed across the experimental site (laser controller, remote alignment units, etc.) and the rest of it is mounted inside the diagnostic cabinet. The process logic is implemented in the three main controllers, see Fig.~\ref{fig:timing}: the diagnostic server, the crate NI PXI Express PXIe-1082 and the chassis NI cRIO-9067. To reduce the electromagnetic pickup on APD signals, all external input and output connections go through optical convertors (all timing and sync lines) or optocouplers (RS232 and USB communication lines, analog inputs from FDS100 photodiodes of the alignment monitor, other). Fig.~\ref{fig:timing} contains the block diagram of the TS timing system.
The timing clock of $10~Hz$ for the laser and data recording system is generated in the Field Programmable Gate Array (FPGA) Xilinx Zync-7000 of the NI cRIO station. It also receives the external sync pulse issued by the GDT control coming at $t_{sync}=-10~s$ relative to the plasma ignition moment $t_{GDT}=0$ and aligns the output clock of $10~Hz$ to fire the laser at the set time point. The internal pattern of timing pulses is provided by the 8-channel digital delay generator GreenField Technologies GFT9404 and repeaters. This versatile FPGA based system is capable to program various timing schemes. The TS data collection system has been already tested with the timing scheme having the maximum measurement repetition frequency of $50~kHz$ for upcoming time-resolved experiments with a pulse-burst laser.
Spectrometers with the integrated signal processing electronics \cite{TS_GDT-spectro} are linked in the `instrument' local area network via the Ethernet-1000 interface, see Fig.\ref{fig:timing}. The software frontend running on the server, is developed using NI LabView and OriginLab OriginPro distributions.

\section{\label{sec:results} Measurements of electron temperature and density}
\subsection{\label{sec:te_meas}Radial profiles of $T_e$}
\begin{figure}[htbp]
\centering 
\includegraphics[width=1.\textwidth]{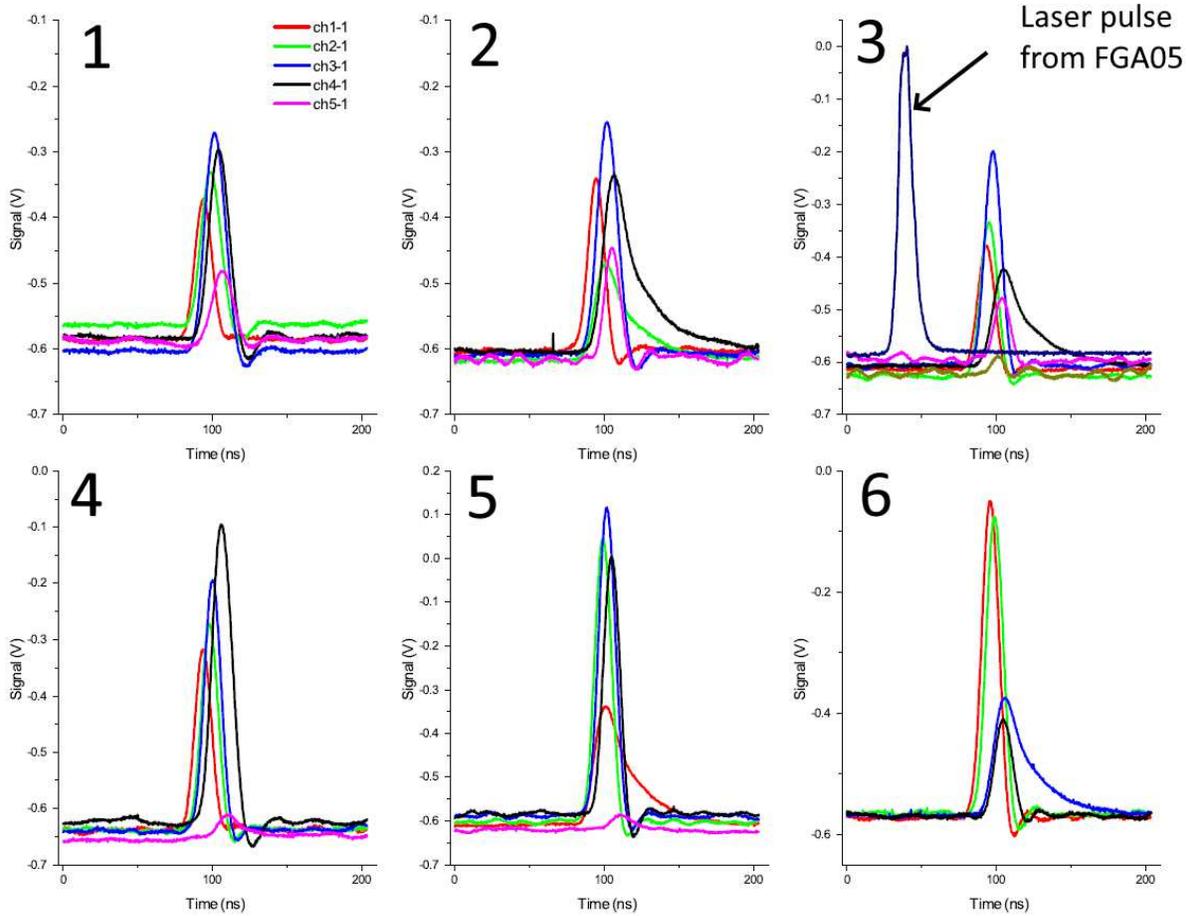}
\caption{\label{fig:signals_ts} TS signals of 6 spectrometers in the shot 51929.}
\end{figure}

In the temperature range $T_e \leqslant 3~keV$, which seems a limitation for observable GDT regimes, the used Sheffield's approximation \cite{sheffield2nd} to the scattering spectrum has a sufficient accuracy of $\leqslant 1\%$ as confirmed by comparison with the Seldon's formula \cite{SELDEN1980405} and numerical modelling. Fig.~\ref{fig:signals_ts} shows the example of TS signals acquired in the GDT shot 51929. One may notice that several APD traces like channel-4 in the spectrometer-2, have a longer trailing front. This appeared to be the feature of several APD Hamamtsu S11519-15 (not amplifier's). At the same time, the signal integration remains accurate since the signal fits the record time window. As it it seen in Fig.~\ref{fig:signals_ts}, all spectrometers delivered signals of a respectable amplitude in channels-1...5 except for the spectrometer-6 bound to the outmost LOS corresponding to $r_6 = 15.6~cm$. The edge plasma is characterized by a low $T_e \leqslant 40~eV$ and only first four spectral channels contribute data. Along with TS signals, there is the reference laser pulse signal from the fast photodiode FGA05 in the channel-7 of the spectrometer-3. The FGA05 output is absolutely calibrated yielding the pulse energy for conversion of the collected number of photons into the electron density. The time-of-flight delay between different spectral channels (and relative to the signal of FGA05 diode, which is located close to the laser) is clearly visible. It is important to add that no signals of stray laser radiation were observed in any spectrometer channels in shots without the plasma.

To calculate the electron temperature, a standard method of a maximum likelihood (ML) is applied \cite{TS_ML}. In order to reduce the calculation time, the first approximation is done with a lookup in the database where synthetic TS signals are stored with the 10~eV step. The following accurate solution engages the iterative ML numerical procedure in the vicinity of found raw estimation.

\begin{figure}[htbp]
\centering 
\includegraphics[width=1.\textwidth]{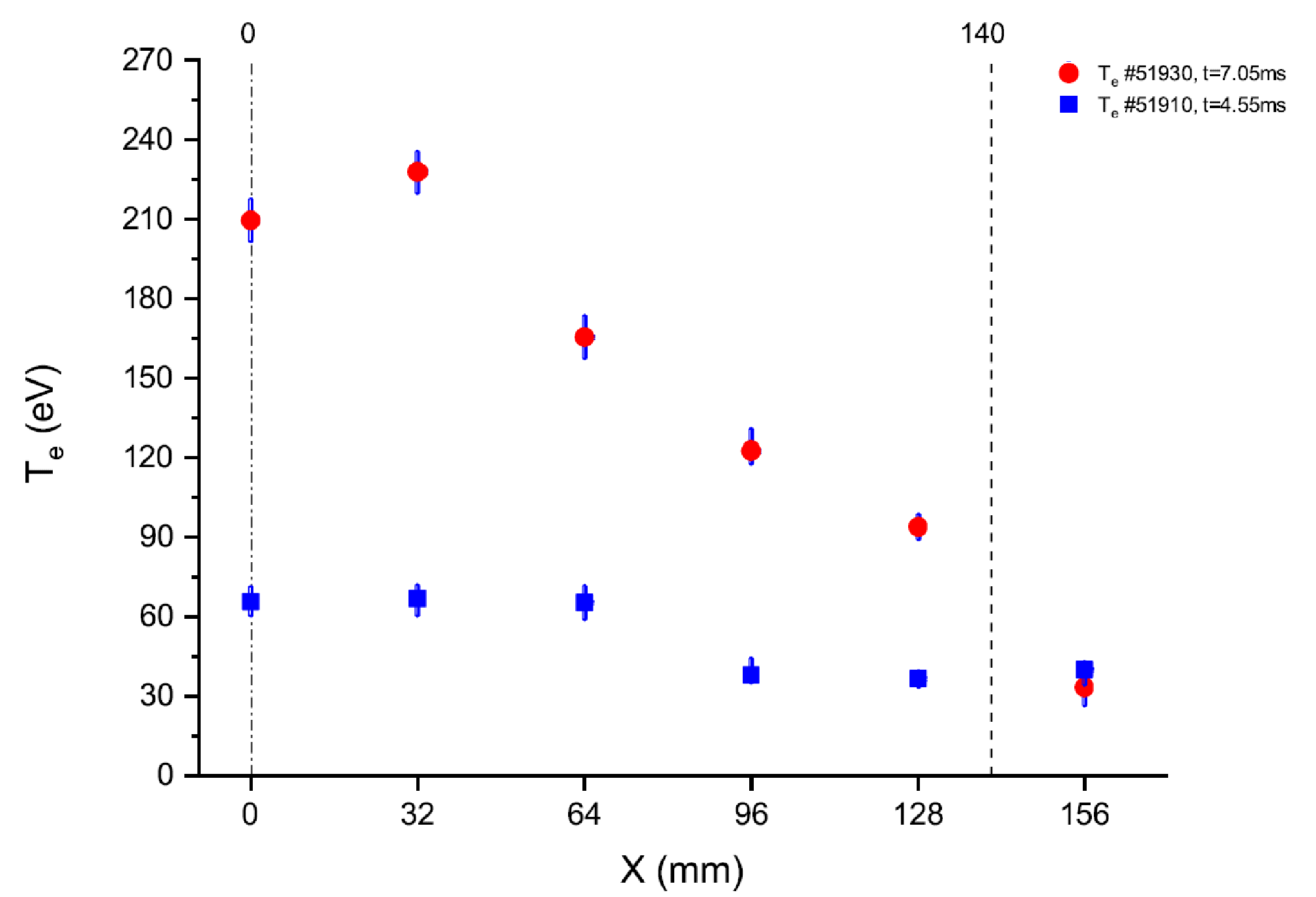}
\caption{\label{fig:te_profiles} Radial profiles of electron temperature.}
\end{figure}  

Two examples of the electron temperature radial profile are plotted in Fig.~\ref{fig:te_profiles}. The data is acquired in two GDT shots: 51910 (blue rectangles) at $t=4.55~ms$ and 51930 (red circles) at $t=7.05~ms$. Dash-dot and dash lines mark the axis and the radial limiter position, respectively. Both measurements correspond to the regime of moderate temperatures without ECR heating. Unlike the recently adopted scenario with the ECR breakdown for plasma startup, the initial target plasma is generated by the on-axis arc discharge source, which operation is always accompanied by an enhanced longitudinal electron heat loss in the core region. This explains the dip near the axis on both profiles.
\subsection{\label{sec:RR_calibr}Intensity calibration by rotational Raman scattering}
\begin{figure}[htbp]
\centering 
\includegraphics[width=1.\textwidth]{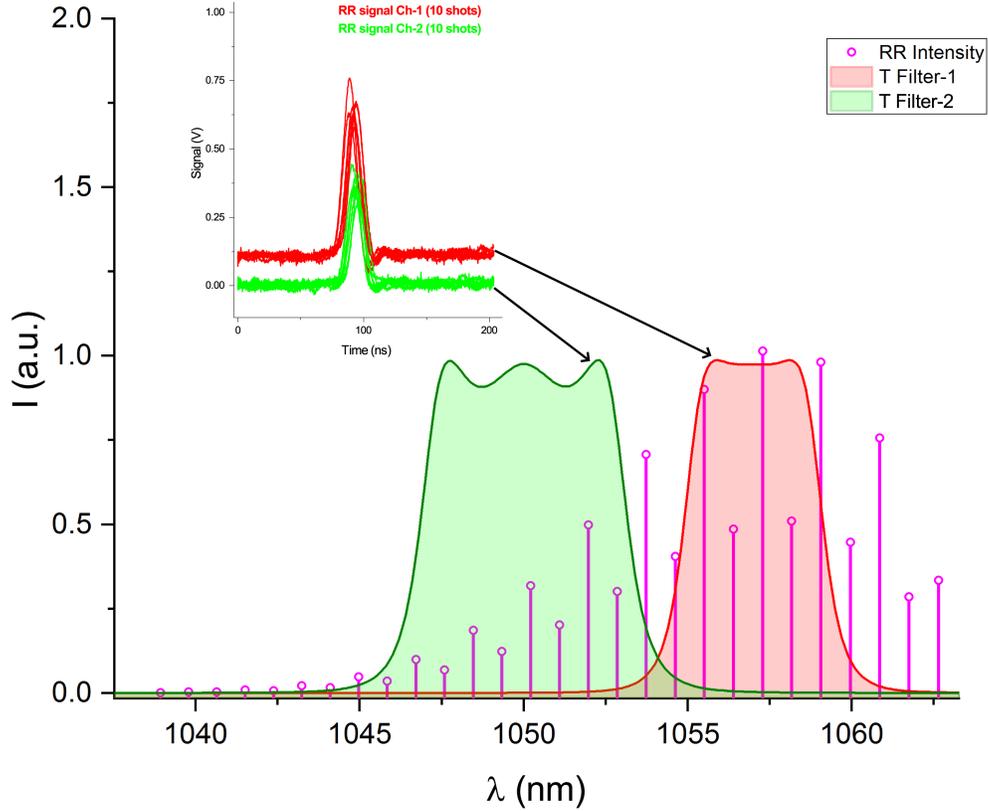}
\caption{\label{fig:RR_calibr} Relative intensities of anti-Stokes RR transitions for $N_2$ and pass band curves for spectral channels-1,2. Insert: accumulated RR signals over 10 shots in channels-1,2.}
\end{figure}

Pass bands of a spectrometer's interference filters \cite{TS_GDT-spectro} restrict using the unshifted Rayleigh scattering to calibrate the system throughput for measurements of the electron density. Measurements at the laser wavelength are inevitably contaminated by a massive stray light making this option additionally unprofitable. For our spectrometer's design, it is convenient to observe anti-Stokes rotational Raman (RR) scattering transitions for $N_2$ gas \cite{TS_RR1,TS_NSTX_RR,TS_W7X_RR}. Equations and numerical data for RR scattering line intensities are taken from \cite{TS_RR_data}. The calibration coefficient for density measurements is defined through relation of TS and RR scattering signals divided by the cross section and the laser pulse energy (see, for example \cite{TS_NSTX_RR}, page 10E737-2). Fig.~\ref{fig:RR_calibr} shows calculated anti-Stokes RR intensities for $N_2$ at the room temperature of $24^\circ~C$ and the pressure of $7.4~mbar$. Overlaid curves represent pass bands of spectrometer's channels-1,2. The given pressure is sufficient to obtain similar signal amplitude $U_{1,2}^{RR} \sim 1~V$ in these two channels and it is low enough to make the procedure swift without waiting for dust in the vacuum chamber to settle down.

The insert in Fig.~\ref{fig:RR_calibr} shows 10 signals accumulated for RR scattering calibration, where the amplitude jitter is due to pulse energy variations within the series. Accordingly, each of two active channels provide an independent calibration data for a TS signal absolute intensity permitting to increase the accuracy.
\subsection{\label{sec:te_meas}Profiles and time evolution of $n_e$ and $T_e$}

\begin{figure}[htbp]
\centering 
\includegraphics[width=1.\textwidth]{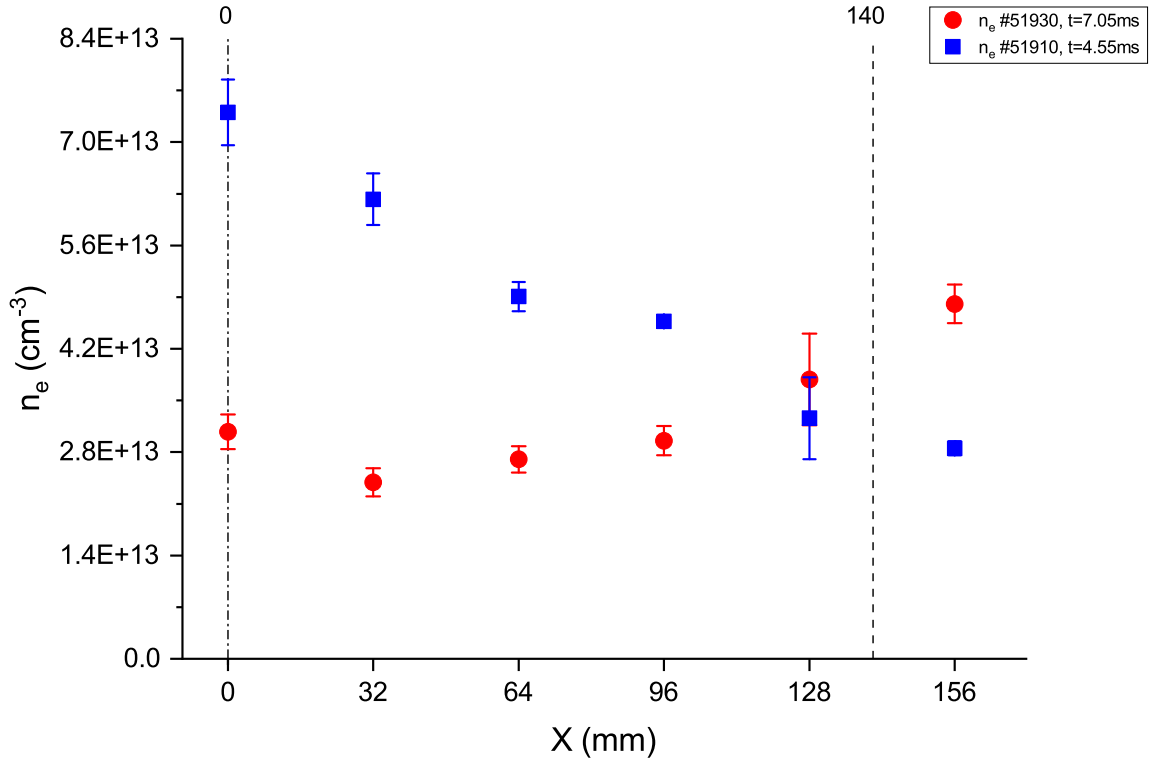}
\caption{\label{fig:ne_profiles} Radial profiles of electron density.}
\end{figure}

\begin{figure}[htbp]
\centering 
\includegraphics[width=1.\textwidth]{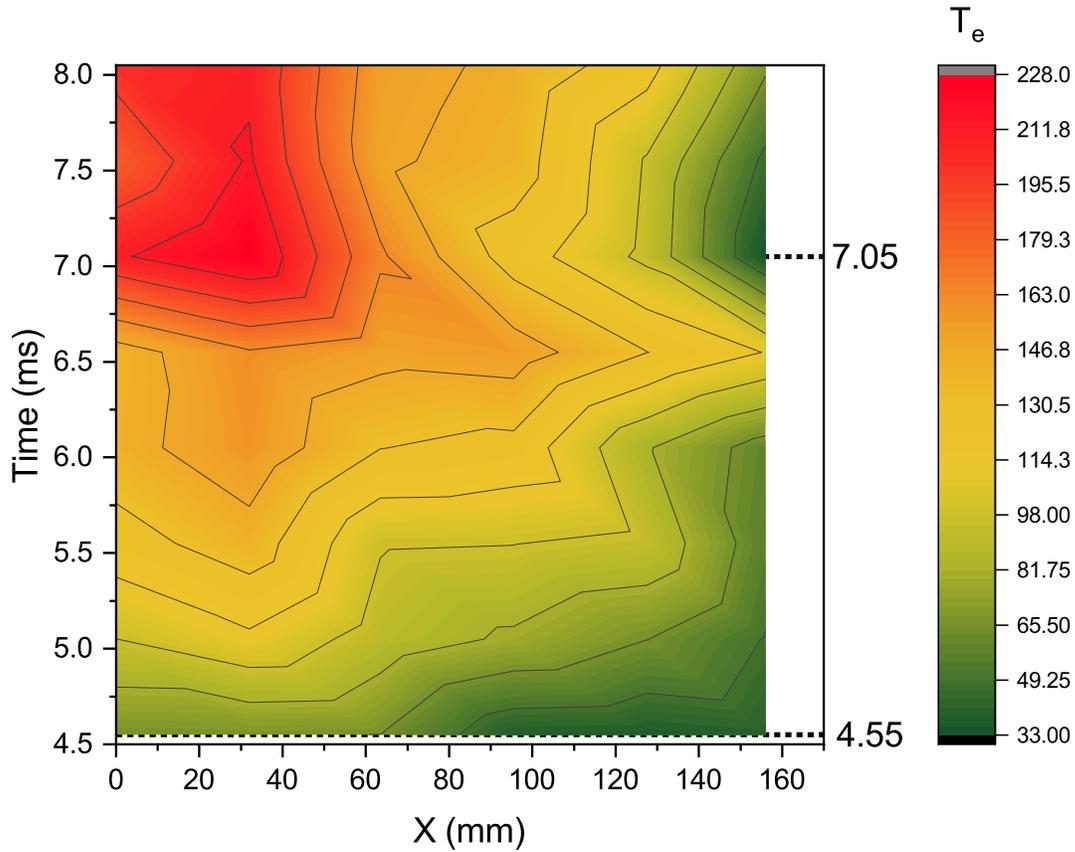}
\caption{\label{fig:te_rt} Time evolution of electron temperature profile.}
\end{figure}

\begin{figure}[htbp]
\centering 
\includegraphics[width=1.\textwidth]{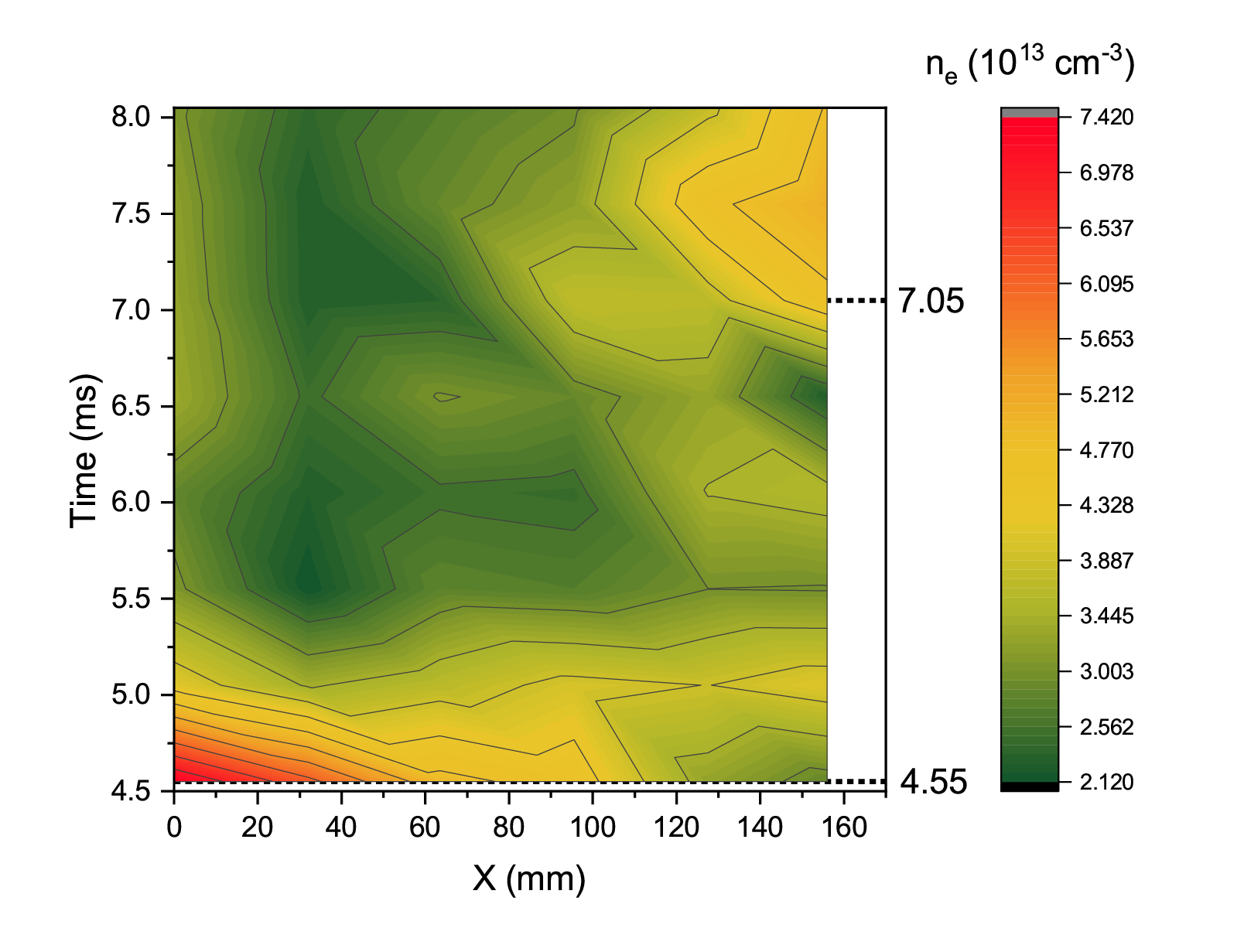}
\caption{\label{fig:ne_rt} Time evolution of electron density profile.}
\end{figure}

The two examples of $n_e(r)$ profiles are shown in Fig.~\ref{fig:ne_profiles} acquired in shots 51910, 51930 at the same times $t_1=4.55~ms, t_2=7.05~ms$ as electron temperature profiles in Fig.~\ref{fig:te_profiles}. During the plasma startup and initial atomic beam injection phase, the density profile remains relatively narrow and centre-peaked. Later on, the periphery gas puff maintaining the plasma particle balance, drives the profile transformation to a more hollow shape as one can see for $t_2=7.05~ms$. In more detail, the time evolution of $T_e(r)$ and $n_e(r)$ is illustrated by Fig.~\ref{fig:te_rt} and Fig.~\ref{fig:ne_rt}. The plots are constituted of data points gathered in several shots so they only reflect a typical time dependence.

\section{\label{sec:summary}Summary and outlook}
Development and commissioning of the Thomson scattering diagnostic at GDT is complete and results of the first experimental campaign involving the new instrument, are presented in the paper. Six spatial points with the resolution of $\delta_r \approx 1~cm$ arranged across the plasma radius of $15.6~cm$, give a satisfactory information on the profile shape. In the regimes with the electron density $n_e \geqslant 10^{19}~m^{-3}$, we have obtained a good measurement accuracy of few percent (RMS error $<5\%$ in the absolute majority of cases) for both the temperature and the density. All spectral channels are completely free from the stray laser radiation thanks both to the interference filters with a high damping ratio $\geqslant 10^5~@1064~nm$\cite{TS_GDT-spectro} and the beamline design. 

Study of dynamics of fast ion and electron distribution functions in GDT plasmas requires time resolved measurements with the Thomson scattering at a repetition rate not less than $1~kHz$, up to $4 \div 5~kHz$ to discover transient events with the ECRH auxiliary heating. Similarly to a successful application of pulse burst lasers in plasma experiments of a sub-second duration \cite{Young_2013,TS_C-2W} and tokamaks \cite{Hartog_2017}, we are planning to install the second laser firing at rates of up to $25~kHz$ in a 10-shot train. The upgrade is scheduled at early 2023.

\nocite{*}
\bibliography{Spectro_LS_lizunov}

\end{document}